\documentclass[twocolumn]{aastex631}

\usepackage{url}
\usepackage{color}
\usepackage{float}
\usepackage{graphicx}

\shorttitle{High-Energy Emission Models with X-ray Polarimetry}
\shortauthors{Peirson et al.}

\graphicspath{{./}{figures/}}

\begin{document}

\title{Testing High-Energy Emission Models for Blazars with X-ray Polarimetry}

\author[0000-0001-6292-1911]{Abel L. Peirson}
\email{alpv95@stanford.edu}
\affiliation{Kavli Insitute for Particle Astrophysics and Cosmology,
Stanford University, Stanford, CA 94305, USA}

\author[0000-0001-9200-4006]{Ioannis Liodakis}
\affiliation{Finnish Centre for Astronomy with ESO, University of Turku, Vesilinnantie 5, FI-20014, Finland}

\author[0000-0001-6711-3286]{Roger W. Romani}
\affiliation{Kavli Insitute for Particle Astrophysics and Cosmology,
Stanford University, Stanford, CA 94305, USA}

\begin{abstract}
Both leptonic and hadronic emission processes may contribute to blazar jet emission; which dominates in blazars's high energy emission component remains an open question. Some intermediate synchrotron peaked blazars transition from their low to high energy emission components in the X-ray band making them excellent laboratories to probe both components simultaneously, and good targets for the newly launched Imaging X-ray Polarimetry Explorer. We characterize the spectral energy distributions for three such blazars: CGRaBS~J0211+1051, TXS~0506+056, and S5~0716+714, predicting their X-ray polarization behavior by fitting a multizone polarized leptonic jet model. We find that a significant detection of electron synchrotron dominated polarization is possible with a 300~ks observation for S5~0716+714 and CGRaBS~J0211+1051 in their flaring states, while even 500~ks observations are unlikely to measure synchrotron self-Compton polarization. Importantly, non-leptonic emission processes like proton synchrotron are marginally detectable for our brightest ISP, S5~0716+714, during a flaring state. Improved {\it IXPE} data reduction methods or next generation telescopes like {\it eXTP} are needed to confidently measure SSC polarization.
\end{abstract}


\section{Introduction} \label{sec:intro}

Blazars are active galactic nuclei whose relativistic jets are oriented at an angle $\theta_{\rm obs}$ within a few degrees, typically $<15^o$ \citep{Liodakis2018}, from an observer on Earth. This results in the relativistically boosted emission from the jet to outshine the host galaxy. The jet's observed multiwavelength emission, from radio to $\gamma$-rays, is characterized by two broad components. The low-energy component is attributed to synchrotron emission from primary jet electrons, while the high-energy component is still unknown with inverse-Compton (IC) scattering or hadronic processes (proton synchrotron, pion cascades, etc.) as the current favored mechanisms \citep{Blandford2019}. Blazars are often classified by the peak frequency ($\nu_{Sy}$) of the low-energy component \citep{Abdo2010}. Here we focus on the ``Intermediate Synchrotron Peak'' (ISP) blazars, whose synchrotron emission  peaks in optical/UV and have their spectral energy distributions (SEDs) dropping toward eventual high-energy component dominance in the hard X-ray/$\gamma$-ray band. In particular, we focus on a subclass of ISPs whose X-ray emission lies in the valley formed by the combination of the two spectral components.

While the origin of the high-energy component is still unknown, the recent launch of the Imaging X-ray Polarimetry Explorer ({\it IXPE}, \citealp{Weisskopf2022}) offers a new diagnostic tool to probe the jet physics, composition, and acceleration of particles. In \cite{Liodakis2019} we used a multizone jet model \citep{Marscher2014,Peirson2018,Peirson2019} and optical polarization results from the RoboPol survey \citep{Blinov2021} to make predictions for the X-ray polarization degree of blazars.  In a synchrotron self-Compton (SSC) scenario, we expect substantial polarization from the electron synchrotron and much lower polarization levels from the Compton component. On the other hand, the polarization degree of proton-synchrotron and synchrotron from hadron initiated pair cascades is expected to be much higher than that of IC emission and comparable to that of the primary electron synchrotron component. Interestingly, as one observes further out on the electron synchrotron cut-off tail, fewer jet emission zones have sufficient particle energies and Doppler factors to produce the detected radiation -- this means that there is decreased polarization angle (PA) averaging between emission zones and thus larger net synchrotron polarization degree and higher variability \citep{Peirson2018,Peirson2019}. Polarization measurements in the transition region between low- and high-energy components can thus be a powerful tool to probe not only for the high-energy emission processes, but also the jet and magnetic field structure. Coincidentally, recent hybrid (also known as lepto-hadronic) blazar models for the high-energy neutrino emission suggest the existence of subdominant synchrotron components from proton initiated pair cascades that might only be detectable in the transition valley where any primary lepton emission is minimized \cite[e.g.,][]{Gao2019}. All of the above suggest that ISPs, whose transition regions lie in the  1-10\,keV band are particularly attractive targets for current and future X-ray polarization missions.

We have identified three such sources, namely CGRaBS~J0211+1051, TXS~0506+056, and S5~0716+714. CGRaBS J0211+1051 and S5~0716+714 are first year {\it IXPE} targets, while  CGRaBS J0211+1051 and TXS 0506+056 are potential neutrino emitters \citep{icecube2018-II,Hovatta2021}. Our goal is twofold: (1) understand the polarization behaviour of the jet across the transition region; (2) make predictions for {\it IXPE} and future missions to understand the high-energy emission signatures from blazars. In section \ref{sec:d&m} we describe the data and jet models, in section \ref{sec:ixpe} we make predictions for {\it IXPE}, and in section \ref{sec:disc} we discuss our findings.

\section{Multiwavelength observations and modeling} \label{sec:d&m}

The non-X-ray multiwavelength data for all sources are taken from the Space Science data center archive\footnote{\url{https://tools.ssdc.asi.it/SED/}}. The data are not contemporaneous and include both flaring and quiescent periods of each source. Since the latter may allow improved polarization measurements, we analyze quiescent and flaring data separately. We bin the observations in frequency bins of 0.1 dex and treat the resulting SED as an ``average'' SED of the source in the given state. Examples are shown in Fig.~\ref{fig:sed}.

For the X-ray observations we used publicly available Neil Gehrels Swift Observatory ({\it Swift}), {\it NuSTAR} and XMM-Newton data from the High Energy Astrophysics Science Archive Research Center (HEASARC) browse interface. The X-ray observations for the source in normal and flaring (f) states are drawn from the following date ranges: 
CGRaBS J0211+1051(f) (MJD 55260-55886),
CGRaBS J0211+1051 (MJD 59250),
S5~0716+714(f) (MJD 57046),
S5~0716+714 (MJD 54864-55920),
TXS 0506+056(f) (MJD 58025).
For CGRaBS J0211+1051 in quiescence we augment with a new $\sim65$~ks XMM-Newton observation (AO-19, ID number: 0861840101, MJD 59250). All {\it Swift} data used were contemporaneous (within 1-2 days) of exposures from one of the other facilities.  The data were processed using the standard HEASARC tools and recommended analyses.

Previous fits to these data have typically used absorbed broken powerlaw models for TXS~0506+056 \citep{icecube2018-II} and S5 0716+714 \citep{Wierzcholska2016}. Instead, we fit the extracted spectra, using {\it XSPEC}, with a more physically motivated model: the sum of two powerlaw components, subject to absorption by a Galactic neutral hydrogen column $\rm N_H$. For CGRaBS J0211+1051, the publicly available {\it Swift} snapshots did not provide sufficient signal-to-noise to unambiguously determine the shape of the 1-10~keV spectrum. We thus tried three models, a single power-law, a power-law with an exponential cut-off, and a sum of two power-laws. We then used the Akaike information criterion to select the model that best describes the data. Again the sum of two power-laws is preferred. The best-fit model parameters are given in Table \ref{tab:xspec}. 

\begin{deluxetable}{lccc}[h]
\tablecaption{X-ray spectral parameters  \label{tab:xspec}}
\tablehead{\colhead{Name} & \colhead{$\rm N_H$} & \colhead{$\Gamma_2$}  & \colhead{$\Gamma_1$}}
\startdata
CGRaBS~J0211+1051 & 0.16 & 1.18$\pm0.19$  & 2.68$\pm$0.4 \\
TXS~0506+056 & 0.25 & 1.83$\pm$0.2 & 3.88$\pm$1.0 \\
S5~0716+714 & 0.031 & 1.56$\pm0.33$ & 2.36$\pm$0.14 \\
    \enddata
 \tablecomments{The $\rm \rm N_H$ values are given in units of $\times10^{22}~cm^{-2}$.}
\end{deluxetable}

\subsection{SED modeling}
Our joint SED and polarization modeling uses the polarized leptonic jet emission model developed in \citet{Peirson2018, Peirson2019}, inspired partly by \citet{potter_synchrotron_2012, Marscher2014}. It assumes an initial power-law electron population propagating along a relativistic conical jet. The jet cross section is divided into multiple magnetic field zones, with isotropically distributed field orientations. These magnetic fields are comoving with the jet material.
Polarized synchrotron emission is self-consistently calculated as the electron population propagates and cools. SSC emission is computed, including the propagation of synchrotron photons from downstream emission into each Comptonizing zone. For quasi-spherical magnetic field zones, as often assumed in turbulent scenarios, the model resolves a decorrelation timescale (which depends on the initial jet parameters) of 0.5--5 days. This is the timescale over which steady state jet emission is expected to fluctuate.

An important feature of this jet model is variable Doppler boosting of the zones, since those directed closest to the line of sight increasingly dominant in the observed flux as the SED steepens \citep{Peirson2019}. This guarantees an increasing expected polarization degree and larger polarization variability above the synchrotron peak. The X-ray SSC polarization degree is typically $\sim 0.2-0.35\times$ that of the synchrotron peak polarization; the components' PAs are highly correlated. In this model the observed synchrotron and SSC polarization behavior depends significantly on the geometric jet parameters, such as jet opening angle, observation angle, and Lorentz factor.

Our leptonic jet emission model is essentially independent of particle acceleration method since it follows zones downstream of any acceleration region. The assumed `chaotic' disordered magnetic fields and relativistic boosting effects should be present in many magnetic reconnection and shock acceleration scenarios. We also assume steady-state emission, probing polarization variability by re-seeding the magnetic field zones. We briefly discuss how these simplifying assumptions may be violated in other models found in the literature in section~\ref{sec:disc}.

\begin{figure}
\centering
\includegraphics[width=0.5\textwidth]{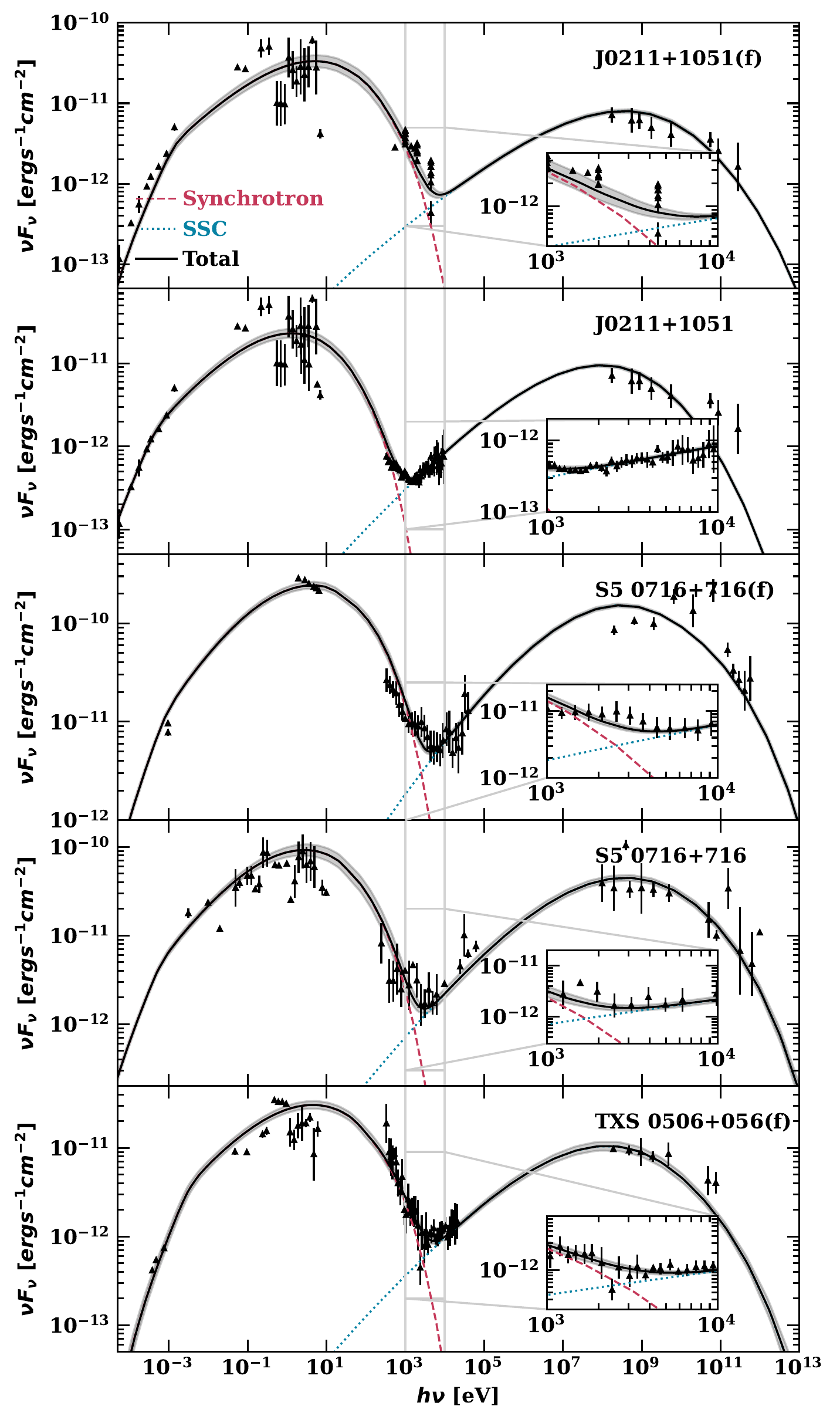}
\caption{Polarized leptonic jet model fits to all blazars and states. `(f)' denotes a flaring state. Black traces show the expected total SED for the best fit jet parameters. Grey shaded regions around the black trace show $1\sigma$ model deviations due to different random magnetic field zones. Vertical grey lines denote {\it IXPE}'s sensitive band, $1-10$~keV, and insets show close-ups of this region.}
\label{fig:sed}
\end{figure}

In order to constrain our model's jet parameters, we fit the multiwavelength SED observations of each blazar state. Due to the chaotic magnetic field zones, our model is stochastic: the same jet parameters can result in different observed SEDs. A stochastic optimization method is necessary to fit such a model to fixed observations. We use a simple variant of the cross-entropy method \citep{rubinstein_cross_2004,kochenderfer_algorithms_2019}. At each step, this samples $n$ sets of jet parameters from a multivariate Gaussian and re-fits a new Gaussian using $k$ samples with the lowest $\chi^2$. Steps are repeated until convergence, when the mean and covariance matrix of the Gaussian no longer change significantly between steps. We use $n=80, k=20$. Since the multiwavelength SEDs for each blazar are not simultaneous and the true SEDs can be much more variable than the observational errors imply, we make the simplifying assumption that every observation has the same error. Our jet model has 8 free parameters. We open source the code to run our model and reproduce the results\footnote{\texttt{https://github.com/alpv95/SSCpol}} \citep{peirson_alpv95sscpol_2022}.

\begin{figure}
    \centering
    \includegraphics[width=0.5\textwidth]{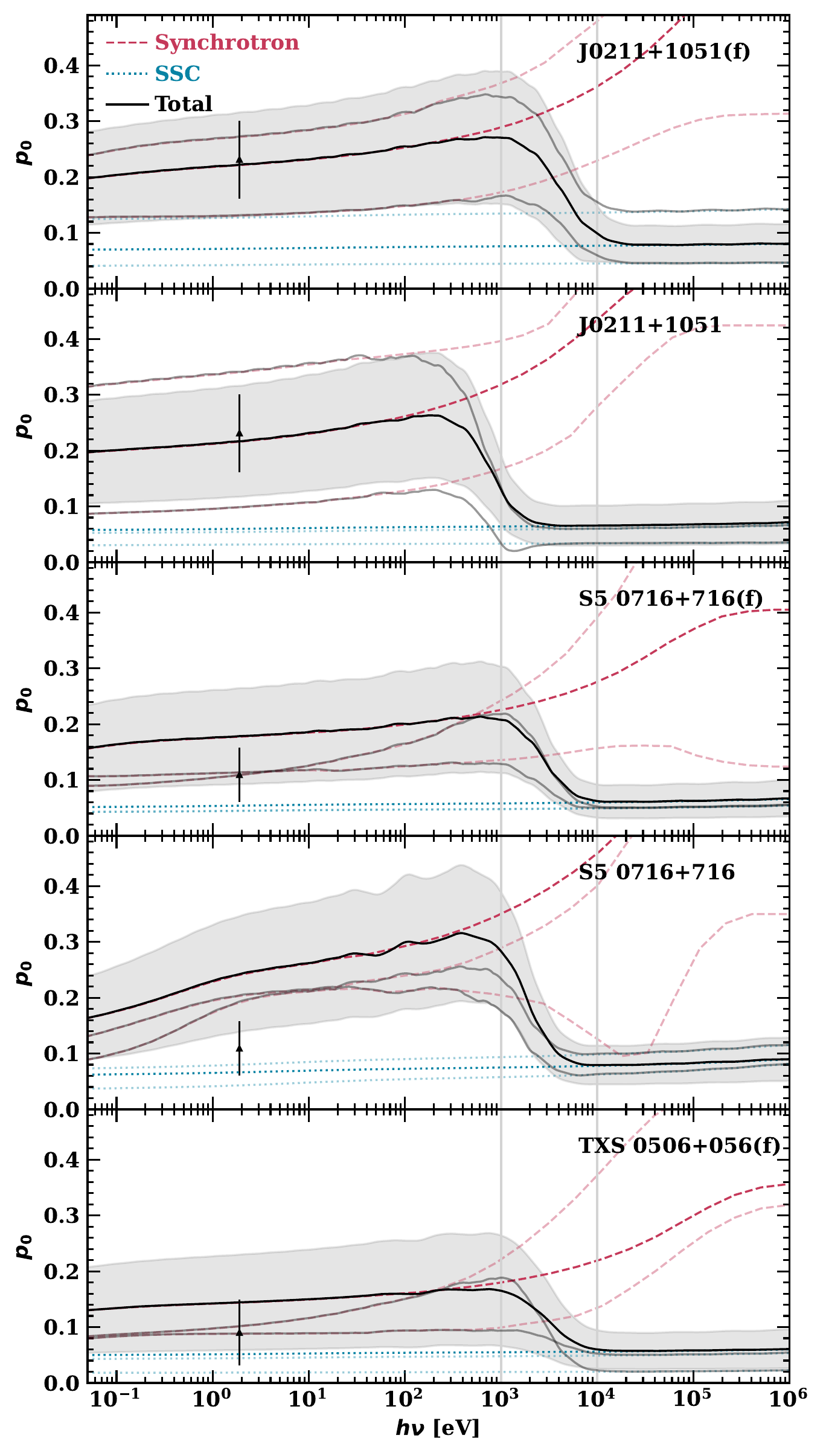}
    \caption{Leptonic (SSC) jet model polarization fraction predictions. The jet models used in each panel are the same as those in Fig.~\ref{fig:sed}. Black observations denote the average measured optical polarization over multiple epochs (MJD 56432 -- 57893 for Robopol measured S5~0716+714 and J0211+1051 \citep{Blinov2021} and MJD 58019 -- 58267 for TXS~0506+056). Lines and shaded regions mean the same as Fig.~\ref{fig:sed} with the addition of two transparent models, which represent randomly selected model realizations.}
    \label{fig:pol}
\end{figure}

Model fit results are shown in Fig.~\ref{fig:sed}. Best fit jet parameters and their respective errors are displayed in Table~\ref{tab:jet}. In Fig.~\ref{fig:pol} we show the predicted polarization behavior resulting from the jet model fits displayed in Fig.~\ref{fig:sed}. The number of magnetic field zones in the jet model is selected so that the predicted optical polarization fraction matches the median of the observations \citep{Blinov2021} as closely as possible. Note that individual realizations of the polarization fraction can vary significantly.

It is useful to compare our polarization predictions to previous studies. \citet{zhang_probing_2019} model TXS 0506+056 using a single zone leptonic emission model with a uniform magnetic field, matching the observed optical polarization degree with a constant polarization dilution factor. They predict an SSC polarization degree of approximately $5\%$ in the X-ray band, rising to $8\%$ at MeV energies. This represents a slightly higher X-ray polariztion fraction, increasing more strongly to high energy. The differences can be attributed to our multizone setup, where multiple magnetic field orientations relative to the line of sight affect the net synchrotron to SSC polarization ratio and its energy dependence \citep{bonometto_polarization_1973, Peirson2019}; multi-zone models generally predict lower SSC polarization degree. We note that our model also propagates synchrotron seed photons between magnetic field zones, further diluting the SSC polarization and increasing sensitivity to the jet geometry.

\begin{deluxetable*}{lccccccccc}[th]
\tablecaption{Polarized jet model best fit parameters. Jet power $W_j$, electron high energy cutoff before exponential decay $E_{\rm max}$, electron population power law index $\alpha$, full conical jet opening angle in lab frame $\theta_{\rm open}$, bulk Lorentz factor $\Gamma_{\rm bulk}$, initial magnetic field strength $B_0$, jet observation angle in lab frame $\theta_{\rm obs}$, and initial equipartition fraction $A_{\rm eq}$. \label{tab:jet}}
\tablehead{\colhead{Name} &\colhead{$W_j [10^{37}W]$} & \colhead{$E_{\rm max} [10^{9} eV]$} & \colhead{$\alpha$}  & \colhead{$\theta_{\rm open} [^\circ]$} & \colhead{$\Gamma_{\rm bulk}$} & \colhead{$B_0 [10^{-5}T]$} & \colhead{$\theta_{\rm obs} [^\circ]$}  & \colhead{$A_{\rm eq}$}}
\startdata
J0211+1051(f) & $4.94 \pm 0.1$ & $13.4 \pm 1.0$ & $2.05 \pm 0.01 $ & $7.16 \pm 0.19$ & $14.8\pm 0.64$ & $5.04 \pm 0.2$ & $1.95\pm 0.18$ & $0.81\pm 0.01$\\
J0211+1051 & $6.48 \pm 0.2$ & $9.59 \pm 0.1$ & $1.85 \pm 0.01 $ & $14.1 \pm 0.35$ & $7.23\pm 0.15$ & $2.38 \pm 0.2$ & $2.31\pm 0.09$ & $0.83\pm 0.02$\\
TXS~0506+056 & $5.26 \pm 0.5$ & $8.03 \pm 0.3$ & $1.89 \pm 0.01 $ & $4.15 \pm 0.26$ & $17.3\pm 0.42$ & $9.63 \pm 0.3$ & $1.64\pm 0.11$ & $0.98\pm 0.05$\\
S5~0716+714(f) & $42.4 \pm 5.0$ & $13.4 \pm 3.0$ & $1.66 \pm 0.02 $ & $7.35 \pm 0.62$ & $13.2\pm 0.51$ & $2.84 \pm 0.6$ & $2.51\pm 0.06$ & $1.05\pm 0.02$\\
S5~0716+714 & $47.3 \pm 11.0$ & $9.49 \pm 1.5$ & $1.75 \pm 0.09 $ & $5.93 \pm 0.84$ & $17.0\pm 0.64$ & $3.23 \pm 1.5$ & $4.48\pm 0.16$ & $0.84\pm 0.13$\\
    \enddata
 \tablecomments{(f) denotes a flaring state. The number of magnetic field zones $N_{\rm zones}$ is selected from $[1,7,19,37,64,128]$. All blazar models shown here use 37 magnetic field zones except CGRaBS J0211+1051, which uses 19. }
\end{deluxetable*}

\section{{\it IXPE} Measurement Simulations}\label{sec:ixpe}

A principal goal of {\it IXPE}-ISP source measurements is to detect two different X-ray polarizations -- a lower energy, electron synchrotron dominated component and a higher energy component.  Assuming an SSC spectrum, we explore whether such a measurement is possible for each of our ISPs with typical {\it IXPE} exposures, using {\it IXPE}'s standard analysis pipeline processing over a 2-8~keV band.  

\begin{figure}
    \centering
    \includegraphics[width=0.5\textwidth]{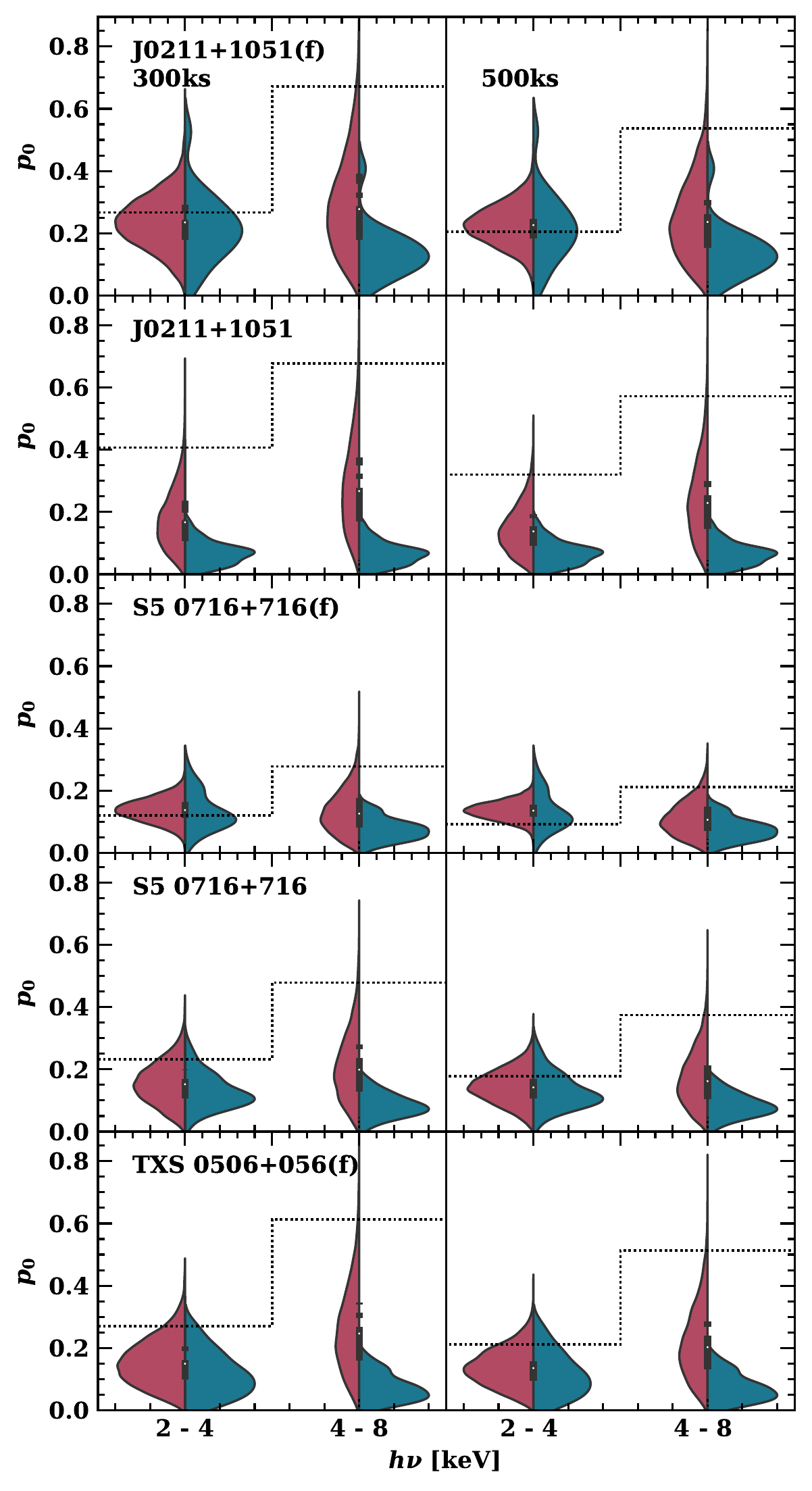}
    \caption{Violin plots of the true polarization fraction distribution (blue, left-hand-side) and the measured polarization fraction distribution (red, right-hand-side) for $2-4$\,keV and $4-8$\,keV energy bins, and 300\,ks/500\,ks exposures. The distributions of true polarization fractions are extracted from our jet model fits (Fig.~\ref{fig:pol}) and are the same for both exposure times. Measured polarization fraction distributions assume a single observation with true polarization equal to the expected value.
    Dashed black traces represent the minimum detectable polarization (MDP$_{99}$) for each measurement bin. }
    \label{fig:meas}
\end{figure}

Using \textit{ixpeobssim}, {\it IXPE}'s observation simulation software \citep{pesce-rollins_observation-simulation_2019}, we simulate multiple 300\,ks and 500\,ks observations for each blazar state assuming polarization and flux are fixed to their expected (average) values (i.e. the black traces in Figs.~\ref{fig:sed}, \ref{fig:pol}). We split the simulated 2-8\,keV data into two energy bins: 2-4\,keV and 4-8\,keV, extracting the polarization fractions by estimating the Stokes' parameters as in \cite{kislat_analyzing_2015}. Figure~\ref{fig:meas} summarizes the results.

In Fig.~\ref{fig:meas}, energy bins where the true polarization fraction distribution (blue, right-hand-side) is fully below the minimum detectable polarization (MDP$_{99}$) threshold (dotted lines) cannot produce significant ($\gtrsim 3\sigma$) detection of non-zero polarization in the given exposure time. MDP$_{99}$ is the 99th percentile upper confidence bound on polarization fraction for an unpolarized source. Energy bins with some or all of the true polarization distribution above MDP$_{99}$ can have significant detections, if their actual polarization is in the upper portion of the predicted range -- the measurement errors would be approximately given by the measured polarization fraction distributions for the most probable $p_0$ (red, left-hand-side). Planned observation times for first-year {\it IXPE} ISP targets, including CGRaBS~J0211+1051 and S5~0716+714, are expected to range from 200\,ks -- 400\,ks.  

For each blazar and state the two energy bins, 2-4\,keV and 4-8\,keV, contain different relative synchrotron and SSC contributions. Insets in Fig.~\ref{fig:sed} give the relative contributions. In non-flaring states, both energy bins are almost entirely dominated by SSC emission so measurement of the synchrotron cutoff component will not be possible.

Low significance polarization fraction measurements, below MDP$_{99}$, are strongly biased away from $p_0 = 0$. Strict non-negativity of $p_0$ forces measurement posteriors (red, Fig.~\ref{fig:meas}) to be asymmetric and for $\mathbb{E}(\hat{p}_0) > p_0$ (see, esp. CGRaBS~J0211+1051 quiescent panel). This highlights the danger of making polarization inferences using low significance point estimates. The measurement bias can be corrected using appropriate $p_0$ estimators \citep{simmons_point_1985}. 

\section{Discussion} \label{sec:disc}
Under a purely leptonic (SSC) jet model for ISPs, we find that simultaneously detecting significant X-ray polarization from both emission components with a $\leq500$\,ks {\it IXPE} observation is impossible, even considering high $p_0$ fluctuations (see Fig.\ref{fig:meas}). 
For the assumptions used here, a $2.5$\,Ms exposure would be required to measure the median predicted SSC polarization in our brightest source, S5~0716+714, during its high state.
Unfortunately, blazar polarization variability may preclude such long observation times. Optical polarization measurements \citep{Blinov2021} suggest that blazar polarization fraction and PA can vary significantly over time periods $<500$\,ks. This would result in an incoherent averaging of polarization vectors leading to depolarization. Many blazar models \citep{Marscher2014} including our own (Fig.~\ref{fig:pol}) predict polarized X-ray electron synchrotron emission to be more variable than the optical \citep{di_gesu_testing_2022}.

If external Compton (EC) contributes significantly to a blazar's high energy emission component, the case for measuring its X-ray polarization becomes even more dim. EC emission is usually assumed to be unpolarized \citep{zhang_x-ray_2013} since the external photon field being scattered is assumed incoherent, originating in the broad line region or accretion disk. Even a small EC contribution can make observations more difficult because MDP$_{99} \propto 1/\sqrt{N_{\rm ph}}$. A $10\%$ fractional EC contribution would lower Fig.~\ref{fig:meas} true polarization fractions by $10\%$ and increase required observation times for the same significance by $23\%$. 
Luckily, all three sources considered are classified as BL-Lac objects, typically associated with low EC contributions. \citet{padovani_txs_2019}, however, suggest that TXS 0506+056 is an FSRQ in disguise, in which case there might be significant EC contribution to the high energy component.

For the ISPs in flaring states, a significant polarization measurement of the synchrotron cutoff is feasible although still difficult, requiring the blazar to be in a high polarization state. Along the primary synchrotron cutoff we expect an increased expected polarization fraction and variability compared to the optical SED peak (see Fig.~\ref{fig:pol}) as the most Doppler-boosted magnetic field zones increasingly dominate the observed emission \citep{Peirson2019}.
Our model presents the minimal (geometry-induced) increase polarization degree above the primary synchrotron peak; other effects may further increase the dominance of individual zones.
For example, in the shock scenario particles are more efficiently accelerated when the magnetic field is aligned along the shock normal \citep{Marscher2014}. Thus chaotic magnetic field zones will vary in their upper electron energy cutoff and that energy can correlate with the global jet geometry. The highest energy electrons contributing X-ray emission are close to the shock, where cooling is limited and the field orientation (and hence polarization) are more highly correlated. Also, \citet{Tavecchio2018} show that immediately downstream of a shock magnetic field compression increases the field perpendicular to the shock normal; this preferential alignment tends to correlate the field orientations and increase net polarization degree, although such correlation decays as turbulence develops downstream. Both these effects may increase polarization at the high energy end of the synchrotron component, improving measurement prospects.  In contrast magnetic reconnection scenarios suggest synchrotron cutoff polarization with higher variability but similar net polarization degree to the synchrotron peak emission \citep{zhang_radiation_2020, Tavecchio2018}. Thus comparing X-ray polarization to simultaneous optical polarization degree may be able to distinguish these acceleration scenarios.


Unexpectedly large high energy component polarization arising from non-leptonic jet emission is possible and potentially detectable. Our leptonic (SSC) jet model predicts any high energy component polarization should typically be $0.2-0.35 \times$ lower than at the SED optical peak \citep{Peirson2019} with the decrement sensitive to the jet geometry. Of course, this ratio is variable and can occasionally fluctuate to large values $>0.5$, especially if the peak polarization is low, so only multi-epoch trends or long-term averages have predictive power.
In the most optimistic hadronic jet scenario, proton and secondary electron synchrotron dominate the high energy emission component \citep{zhang_x-ray_2013, Gao2019}. High energy component X-ray band polarization fractions would be similar to the SED optical peak (Fig.~\ref{fig:pol}), corresponding to a X-ray/optical polarization ratio of 1, extremely unlikely in a SSC dominant leptonic emission model. Although this would provide a much needed polarization fraction boost, a two component detection would remain out of reach, even for a $\leq500$\,ks {\it IXPE} observation, for all ISPs except S5~0716+714 in its flaring state. 
Indeed, S5~0716+714 is the first ISP {\it IXPE} target, planned for a 300\,ks observation on 31st March 2022; a significant polarization detection for both high and low energy components would be a promising indication of non-leptonic jet emission.

If initial {\it IXPE} observations do not detect significant polarization from either emission component in any ISPs, it will be difficult to rule out non-leptonic processes. Upper polarization fraction limits based on the MDP$_{99}$s in Fig.~\ref{fig:meas} will be too high to make any useful inference about the polarization ratio of the two emission components, even with a strong synchrotron detection at 2-4~keV. However, the measurements' sensitivity may be improved. Bayesian neural network analysis of {\it IXPE} data \citep{peirson_deep_2021, peirson_towards_2021} has been shown to reduce MDP$_{99}$s by up to $25\%$ compared to the standard {\it IXPE} analysis pipeline, as well as increasing {\it IXPE}'s effective energy band to 1-10\,keV. We may also tune the energy range of the `low' and `high' energy detection windows for an individual source's SED, improving our ability to measure or bound the two components' $p_0$. Such improved analysis could, for example, make a flaring 4-8~keV S5~0716+714 SSC polarization detection possible with a 500\,ks observation -- Fig.~\ref{fig:meas}. Although neural network analysis is not yet in production for {\it IXPE}, a re-analysis of borderline observations could reveal missed discoveries. Looking further ahead, the effective area of future X-ray polarization mission {\it eXTP} \citep{zhang_enhanced_2018} should be four times larger than {\it IXPE}'s, reducing MDP$_{99}$s by a factor of $0.5$ \citep{di_gesu_prospects_2020}. Including both improvements, simultaneous measurement of both ISP emission components with a 500~ks observation is well in scope for all the ISPs considered here.

\begin{acknowledgments}

The authors thank the anonymous referee and Dinesh Kandel for useful comments and discussions that helped improved this work. This research has made use of data and/or software provided by the High Energy Astrophysics Science Archive Research Center (HEASARC), which is a service of the Astrophysics Science Division at NASA/GSFC. Based on observations obtained with XMM-Newton, an ESA science mission with instruments and contributions directly funded by ESA Member States and NASA. This research has made use of data from the RoboPol programme, a collaboration between Caltech, the University of Crete, IA-FORTH, IUCAA, the MPIfR, and the Nicolaus Copernicus University, which was conducted at Skinakas Observatory in Crete, Greece. ALP is supported by the Stanford Data Science Scholars Program, NASA
FINESST program (grant 80NSSC19K1407) and grant
NNM17AA26C from the Marshall Space Flight Center.

\end{acknowledgments}

\facilities{{\it IXPE},  {\it NuSTAR}, RoboPol, {\it Swift}, {\it XMM-Newton}}

\software{{\it ixpeobbsim}, {\it SSCpol}}

\bibliography{sample631}

\begin{thebibliography}{}
\expandafter\ifx\csname natexlab\endcsname\relax\def\natexlab#1{#1}\fi
\providecommand{\url}[1]{\href{#1}{#1}}
\providecommand{\dodoi}[1]{doi:~\href{http://doi.org/#1}{\nolinkurl{#1}}}
\providecommand{\doeprint}[1]{\href{http://ascl.net/#1}{\nolinkurl{http://ascl.net/#1}}}
\providecommand{\doarXiv}[1]{\href{https://arxiv.org/abs/#1}{\nolinkurl{https://arxiv.org/abs/#1}}}

\bibitem[{{Abdo} {et~al.}(2010){Abdo}, {Ackermann}, {Agudo}, {Ajello}, {Aller},
  {Aller}, {Angelakis}, {Arkharov}, {Axelsson}, {Bach}, {Baldini}, {Ballet},
  {Barbiellini}, {Bastieri}, {Baughman}, {Bechtol}, {Bellazzini}, {Benitez},
  {Berdyugin}, {Berenji}, {Blandford}, {Bloom}, {Boettcher}, {Bonamente},
  {Borgland}, {Bregeon}, {Brez}, {Brigida}, {Bruel}, {Burnett}, {Burrows},
  {Buson}, {Caliandro}, {Calzoletti}, {Cameron}, {Capalbi}, {Caraveo},
  {Carosati}, {Casandjian}, {Cavazzuti}, {Cecchi}, {{\c{C}}elik}, {Charles},
  {Chaty}, {Chekhtman}, {Chen}, {Chiang}, {Chincarini}, {Ciprini}, {Claus},
  {Cohen-Tanugi}, {Colafrancesco}, {Cominsky}, {Conrad}, {Costamante},
  {Cutini}, {D'ammando}, {Deitrick}, {D'Elia}, {Dermer}, {de Angelis}, {de
  Palma}, {Digel}, {Donnarumma}, {Silva}, {Drell}, {Dubois}, {Dultzin},
  {Dumora}, {Falcone}, {Farnier}, {Favuzzi}, {Fegan}, {Focke}, {Forn{\'e}},
  {Fortin}, {Frailis}, {Fuhrmann}, {Fukazawa}, {Funk}, {Fusco}, {G{\'o}mez},
  {Gargano}, {Gasparrini}, {Gehrels}, {Germani}, {Giebels}, {Giglietto},
  {Giommi}, {Giordano}, {Giuliani}, {Glanzman}, {Godfrey}, {Grenier},
  {Gronwall}, {Grove}, {Guillemot}, {Guiriec}, {Gurwell}, {Hadasch},
  {Hanabata}, {Harding}, {Hayashida}, {Hays}, {Healey}, {Heidt}, {Hiriart},
  {Horan}, {Hoversten}, {Hughes}, {Itoh}, {Jackson}, {J{\'o}hannesson},
  {Johnson}, {Johnson}, {Jorstad}, {Kadler}, {Kamae}, {Katagiri}, {Kataoka},
  {Kawai}, {Kennea}, {Kerr}, {Kimeridze}, {Kn{\"o}dlseder}, {Kocian},
  {Kopatskaya}, {Koptelova}, {Konstantinova}, {Kovalev}, {Kovalev},
  {Kurtanidze}, {Kuss}, {Lande}, {Larionov}, {Latronico}, {Leto}, {Lindfors},
  {Longo}, {Loparco}, {Lott}, {Lovellette}, {Lubrano}, {Madejski}, {Makeev},
  {Marchegiani}, {Marscher}, {Marshall}, {Max-Moerbeck}, {Mazziotta},
  {McConville}, {McEnery}, {Meurer}, {Michelson}, {Mitthumsiri}, {Mizuno},
  {Moiseev}, {Monte}, {Monzani}, {Morselli}, {Moskalenko}, {Murgia},
  {Nestoras}, {Nilsson}, {Nizhelsky}, {Nolan}, {Norris}, {Nuss}, {Ohsugi},
  {Ojha}, {Omodei}, {Orlando}, {Ormes}, {Osborne}, {Ozaki}, {Pacciani},
  {Padovani}, {Pagani}, {Page}, {Paneque}, {Panetta}, {Parent}, {Pasanen},
  {Pavlidou}, {Pelassa}, {Pepe}, {Perri}, {Pesce-Rollins}, {Piranomonte},
  {Piron}, {Pittori}, {Porter}, {Puccetti}, {Rahoui}, {Rain{\`o}}, {Raiteri},
  {Rando}, {Razzano}, {Reimer}, {Reimer}, {Reposeur}, {Richards}, {Ritz},
  {Rochester}, {Rodriguez}, {Romani}, {Ros}, {Roth}, {Roustazadeh}, {Ryde},
  {Sadrozinski}, {Sadun}, {Sanchez}, {Sander}, {Saz Parkinson}, {Scargle},
  {Sellerholm}, {Sgr{\`o}}, {Shaw}, {Sigua}, {Siskind}, {Smith}, {Smith},
  {Spandre}, {Spinelli}, {Starck}, {Stevenson}, {Stratta}, {Strickman},
  {Suson}, {Tajima}, {Takahashi}, {Takahashi}, {Takalo}, {Tanaka}, {Thayer},
  {Thayer}, {Thompson}, {Tibaldo}, {Torres}, {Tosti}, {Tramacere}, {Uchiyama},
  {Usher}, {Vasileiou}, {Verrecchia}, {Vilchez}, {Villata}, {Vitale}, {Waite},
  {Wang}, {Winer}, {Wood}, {Ylinen}, {Zensus}, {Zhekanis}, \&
  {Ziegler}}]{Abdo2010}
{Abdo}, A.~A., {Ackermann}, M., {Agudo}, I., {et~al.} 2010, \apj, 716, 30,
  \dodoi{10.1088/0004-637X/716/1/30}

\bibitem[{{Blandford} {et~al.}(2019){Blandford}, {Meier}, \&
  {Readhead}}]{Blandford2019}
{Blandford}, R., {Meier}, D., \& {Readhead}, A. 2019, \araa, 57, 467,
  \dodoi{10.1146/annurev-astro-081817-051948}

\bibitem[{{Blinov} {et~al.}(2021){Blinov}, {Kiehlmann}, {Pavlidou},
  {Panopoulou}, {Skalidis}, {Angelakis}, {Casadio}, {Einoder}, {Hovatta},
  {Kokolakis}, {Kougentakis}, {Kus}, {Kylafis}, {Kyritsis}, {Lalakos},
  {Liodakis}, {Maharana}, {Makrydopoulou}, {Mandarakas}, {Maragkakis},
  {Myserlis}, {Papadakis}, {Paterakis}, {Pearson}, {Ramaprakash}, {Readhead},
  {Reig}, {S{\l}owikowska}, {Tassis}, {Xexakis}, {{\.Z}ejmo}, \&
  {Zensus}}]{Blinov2021}
{Blinov}, D., {Kiehlmann}, S., {Pavlidou}, V., {et~al.} 2021, \mnras, 501,
  3715, \dodoi{10.1093/mnras/staa3777}

\bibitem[{Bonometto \& Saggion(1973)}]{bonometto_polarization_1973}
Bonometto, S., \& Saggion, A. 1973, Astronomy and Astrophysics, 23, 9.
\newblock \url{http://adsabs.harvard.edu/abs/1973A%26A....23....9B}

\bibitem[{Di~Gesu {et~al.}(2020)Di~Gesu, Ferrazzoli, Donnarumma, Soffitta,
  Costa, Muleri, Pesce-Rollins, \& Marin}]{di_gesu_prospects_2020}
Di~Gesu, L., Ferrazzoli, R., Donnarumma, I., {et~al.} 2020, Astronomy and
  Astrophysics, 643, A52, \dodoi{10.1051/0004-6361/202037886}

\bibitem[{Di~Gesu {et~al.}(2022)Di~Gesu, Tavecchio, Donnarumma, Marscher,
  Pesce-Rollins, \& Landoni}]{di_gesu_testing_2022}
Di~Gesu, L., Tavecchio, F., Donnarumma, I., {et~al.} 2022, Testing particle
  acceleration models for {BL} {LAC} jets with the {Imaging} {X}-ray
  {Polarimetry} {Explorer}, Tech. rep.
\newblock \url{https://ui.adsabs.harvard.edu/abs/2022arXiv220109597D}

\bibitem[{{Gao} {et~al.}(2019){Gao}, {Fedynitch}, {Winter}, \&
  {Pohl}}]{Gao2019}
{Gao}, S., {Fedynitch}, A., {Winter}, W., \& {Pohl}, M. 2019, Nature Astronomy,
  3, 88, \dodoi{10.1038/s41550-018-0610-1}

\bibitem[{{Hovatta} {et~al.}(2021){Hovatta}, {Lindfors}, {Kiehlmann},
  {Max-Moerbeck}, {Hodges}, {Liodakis}, {L{\"a}hteem{\"a}ki}, {Pearson},
  {Readhead}, {Reeves}, {Suutarinen}, {Tammi}, \& {Tornikoski}}]{Hovatta2021}
{Hovatta}, T., {Lindfors}, E., {Kiehlmann}, S., {et~al.} 2021, \aap, 650, A83,
  \dodoi{10.1051/0004-6361/202039481}

\bibitem[{{IceCube Collaboration} {et~al.}(2018){IceCube Collaboration},
  {Aartsen}, {Ackermann}, {Adams}, {Aguilar}, {Ahlers}, {Ahrens}, {Al Samarai},
  {Altmann}, {Andeen}, {Anderson}, {Ansseau}, {Anton}, {Arg{\"u}elles},
  {Auffenberg}, {Axani}, {Bagherpour}, {Bai}, {Barron}, {Barwick}, {Baum},
  {Bay}, {Beatty}, {Becker Tjus}, {Becker}, {BenZvi}, {Berley}, {Bernardini},
  {Besson}, {Binder}, {Bindig}, {Blaufuss}, {Blot}, {Bohm}, {B{\"o}rner},
  {Bos}, {B{\"o}ser}, {Botner}, {Bourbeau}, {Bourbeau}, {Bradascio}, {Braun},
  {Brenzke}, {Bretz}, {Bron}, {Brostean-Kaiser}, {Burgman}, {Busse}, {Carver},
  {Cheung}, {Chirkin}, {Christov}, {Clark}, {Classen}, {Coenders}, {Collin},
  {Conrad}, {Coppin}, {Correa}, {Cowen}, {Cross}, {Dave}, {Day}, {de
  Andr{\'e}}, {De Clercq}, {DeLaunay}, {Dembinski}, {De Ridder}, {Desiati}, {de
  Vries}, {de Wasseige}, {de With}, {DeYoung}, {D{\'\i}az-V{\'e}lez}, {di
  Lorenzo}, {Dujmovic}, {Dumm}, {Dunkman}, {Dvorak}, {Eberhardt}, {Ehrhardt},
  {Eichmann}, {Eller}, {Evenson}, {Fahey}, {Fazely}, {Felde}, {Filimonov},
  {Finley}, {Flis}, {Franckowiak}, {Friedman}, {Fritz}, {Gaisser}, {Gallagher},
  {Gerhardt}, {Ghorbani}, {Glauch}, {Gl{\"u}senkamp}, {Goldschmidt},
  {Gonzalez}, {Grant}, {Griffith}, {Haack}, {Hallgren}, {Halzen}, {Hanson},
  {Hebecker}, {Heereman}, {Helbing}, {Hellauer}, {Hickford}, {Hignight},
  {Hill}, {Hoffman}, {Hoffmann}, {Hoinka}, {Hokanson-Fasig}, {Hoshina},
  {Huang}, {Huber}, {Hultqvist}, {H{\"u}nnefeld}, {Hussain}, {In}, {Iovine},
  {Ishihara}, {Jacobi}, {Japaridze}, {Jeong}, {Jero}, {Jones}, {Kalaczynski},
  {Kang}, {Kappes}, {Kappesser}, {Karg}, {Karle}, {Katz}, {Kauer}, {Keivani},
  {Kelley}, {Kheirandish}, {Kim}, {Kim}, {Kintscher}, {Kiryluk}, {Kittler},
  {Klein}, {Koirala}, {Kolanoski}, {K{\"o}pke}, {Kopper}, {Kopper},
  {Koschinsky}, {Koskinen}, {Kowalski}, {Krings}, {Kroll}, {Kr{\"u}ckl},
  {Kunwar}, {Kurahashi}, {Kuwabara}, {Kyriacou}, {Labare}, {Lanfranchi},
  {Larson}, {Lauber}, {Leonard}, {Lesiak-Bzdak}, {Leuermann}, {Liu}, {Lozano
  Mariscal}, {Lu}, {L{\"u}nemann}, {Luszczak}, {Madsen}, {Maggi}, {Mahn},
  {Mancina}, {Maruyama}, {Mase}, {Maunu}, {Meagher}, {Medici}, {Meier},
  {Menne}, {Merino}, {Meures}, {Miarecki}, {Micallef}, {Moment{\'e}},
  {Montaruli}, {Moore}, {Morse}, {Moulai}, {Nahnhauer}, {Nakarmi}, {Naumann},
  {Neer}, {Niederhausen}, {Nowicki}, {Nygren}, {Obertacke Pollmann}, {Olivas},
  {O'Murchadha}, {O'Sullivan}, {Palczewski}, {Pandya}, {Pankova}, {Peiffer},
  {Pepper}, {P{\'e}rez de los Heros}, {Pieloth}, {Pinat}, {Plum}, {Price},
  {Przybylski}, {Raab}, {R{\"a}del}, {Rameez}, {Rauch}, {Rawlins}, {Rea},
  {Reimann}, {Relethford}, {Relich}, {Resconi}, {Rhode}, {Richman},
  {Robertson}, {Rongen}, {Rott}, {Ruhe}, {Ryckbosch}, {Rysewyk}, {Safa},
  {S{\"a}lzer}, {Sanchez Herrera}, {Sandrock}, {Sandroos}, {Santander},
  {Sarkar}, {Sarkar}, {Satalecka}, {Schlunder}, {Schmidt}, {Schneider},
  {Schoenen}, {Sch{\"o}neberg}, {Schumacher}, {Sclafani}, {Seckel},
  {Seunarine}, {Soedingrekso}, {Soldin}, {Song}, {Spiczak}, {Spiering},
  {Stachurska}, {Stamatikos}, {Stanev}, {Stasik}, {Stein}, {Stettner},
  {Steuer}, {Stezelberger}, {Stokstad}, {St{\"o}{\ss}l}, {Strotjohann},
  {Stuttard}, {Sullivan}, {Sutherland}, {Taboada}, {Tatar}, {Tenholt},
  {Ter-Antonyan}, {Terliuk}, {Tilav}, {Toale}, {Tobin}, {Toennis}, {Toscano},
  {Tosi}, {Tselengidou}, {Tung}, {Turcati}, {Turley}, {Ty}, {Unger}, {Usner},
  {Vandenbroucke}, {Van Driessche}, {van Eijk}, {van Eijndhoven}, {Vanheule},
  {van Santen}, {Vogel}, {Vraeghe}, {Walck}, {Wallace}, {Wallraff}, {Wandler},
  {Wandkowsky}, {Waza}, {Weaver}, {Weiss}, {Wendt}, {Werthebach}, {Westerhoff},
  {Whelan}, {Whitehorn}, {Wiebe}, {Wiebusch}, {Wille}, {Williams}, {Wills},
  {Wolf}, {Wood}, {Wood}, {Woschnagg}, {Xu}, {Xu}, {Xu}, {Yanez}, {Yodh},
  {Yoshida}, {Yuan}, {Fermi-LAT Collaboration}, {Abdollahi}, {Ajello},
  {Angioni}, {Baldini}, {Ballet}, {Barbiellini}, {Bastieri}, {Bechtol},
  {Bellazzini}, {Berenji}, {Bissaldi}, {Blandford}, {Bonino}, {Bottacini},
  {Bregeon}, {Bruel}, {Buehler}, {Burnett}, {Burns}, {Buson}, {Cameron},
  {Caputo}, {Caraveo}, {Cavazzuti}, {Charles}, {Chen}, {Cheung}, {Chiang},
  {Chiaro}, {Ciprini}, {Cohen-Tanugi}, {Conrad}, {Costantin}, {Cutini},
  {D'Ammando}, {de Palma}, {Digel}, {Di Lalla}, {Di Mauro}, {Di Venere},
  {Dom{\'\i}nguez}, {Favuzzi}, {Franckowiak}, {Fukazawa}, {Funk}, {Fusco},
  {Gargano}, {Gasparrini}, {Giglietto}, {Giomi}, {Giommi}, {Giordano},
  {Giroletti}, {Glanzman}, {Green}, {Grenier}, {Grondin}, {Guiriec}, {Harding},
  {Hayashida}, {Hays}, {Hewitt}, {Horan}, {J{\'o}hannesson}, {Kadler},
  {Kensei}, {Kocevski}, {Krauss}, {Kreter}, {Kuss}, {La Mura}, {Larsson},
  {Latronico}, {Lemoine-Goumard}, {Li}, {Longo}, {Loparco}, {Lovellette},
  {Lubrano}, {Magill}, {Maldera}, {Malyshev}, {Manfreda}, {Mazziotta},
  {McEnery}, {Meyer}, {Michelson}, {Mizuno}, {Monzani}, {Morselli},
  {Moskalenko}, {Negro}, {Nuss}, {Ojha}, {Omodei}, {Orienti}, {Orlando},
  {Palatiello}, {Paliya}, {Perkins}, {Persic}, {Pesce-Rollins}, {Piron},
  {Porter}, {Principe}, {Rain{\`o}}, {Rando}, {Rani}, {Razzano}, {Razzaque},
  {Reimer}, {Reimer}, {Renault-Tinacci}, {Ritz}, {Rochester}, {Saz Parkinson},
  {Sgr{\`o}}, {Siskind}, {Spandre}, {Spinelli}, {Suson}, {Tajima}, {Takahashi},
  {Tanaka}, {Thayer}, {Thompson}, {Tibaldo}, {Torres}, {Torresi}, {Tosti},
  {Troja}, {Valverde}, {Vianello}, {Vogel}, {Wood}, {Wood}, {Zaharijas}, {MAGIC
  Collaboration}, {Ahnen}, {Ansoldi}, {Antonelli}, {Arcaro}, {Baack},
  {Babi{\'c}}, {Banerjee}, {Bangale}, {Barres de Almeida}, {Barrio}, {Becerra
  Gonz{\'a}lez}, {Bednarek}, {Bernardini}, {Berti}, {Bhattacharyya}, {Biland},
  {Blanch}, {Bonnoli}, {Carosi}, {Carosi}, {Ceribella}, {Chatterjee}, {Colak},
  {Colin}, {Colombo}, {Contreras}, {Cortina}, {Covino}, {Cumani}, {Da Vela},
  {Dazzi}, {De Angelis}, {De Lotto}, {Delfino}, {Delgado}, {Di Pierro},
  {Dom{\'\i}nguez}, {Dominis Prester}, {Dorner}, {Doro}, {Einecke},
  {Elsaesser}, {Fallah Ramazani}, {Fern{\'a}ndez-Barral}, {Fidalgo}, {Foffano},
  {Pfrang}, {Fonseca}, {Font}, {Franceschini}, {Fruck}, {Galindo}, {Gallozzi},
  {Garc{\'\i}a L{\'o}pez}, {Garczarczyk}, {Gaug}, {Giammaria}, {Godinovi{\'c}},
  {Gora}, {Guberman}, {Hadasch}, {Hahn}, {Hassan}, {Hayashida}, {Herrera},
  {Hose}, {Hrupec}, {Inoue}, {Ishio}, {Konno}, {Kubo}, {Kushida}, {Lelas},
  {Lindfors}, {Lombardi}, {Longo}, {L{\'o}pez}, {Maggio}, {Majumdar},
  {Makariev}, {Maneva}, {Manganaro}, {Mannheim}, {Maraschi}, {Mariotti},
  {Mart{\'\i}nez}, {Masuda}, {Mazin}, {Minev}, {M}, {Mirzoyan}, {Moralejo},
  {Moreno}, {Moretti}, {Nagayoshi}, {Neustroev}, {Niedzwiecki}, {Nievas
  Rosillo}, {Nigro}, {Nilsson}, {Ninci}, {Nishijima}, {Noda}, {Nogu{\'e}s},
  {Paiano}, {Palacio}, {Paneque}, {Paoletti}, {Paredes}, {Pedaletti},
  {Peresano}, {Persic}, {Prada Moroni}, {Prandini}, {Puljak}, {Rodriguez
  Garcia}, {Reichardt}, {Rhode}, {Rib{\'o}}, {Rico}, {Righi}, {Rugliancich},
  {Saito}, {Satalecka}, {Schweizer}, {Sitarek}, {{\v{S}}nidaric
  {\textasciiacute}}, {Sobczynska}, {Stamerra}, {Strzys}, {Suri{\'c}},
  {Takahashi}, {Tavecchio}, {Temnikov}, {Terzi{\'c}}, {Teshima},
  {Torres-Alb{\`a}}, {Treves}, {Tsujimoto}, {Vanzo}, {Vazquez Acosta}, {Vovk},
  {Ward}, {Will}, {S}, {Zaric {\textasciiacute}}, {AGILE Team}, {Lucarelli},
  {Tavani}, {Piano}, {Donnarumma}, {Pittori}, {Verrecchia}, {Barbiellini},
  {Bulgarelli}, {Caraveo}, {Cattaneo}, {Colafrancesco}, {Costa}, {Di Cocco},
  {Ferrari}, {Gianotti}, {Giuliani}, {Lipari}, {Mereghetti}, {Morselli},
  {Pacciani}, {Paoletti}, {Parmiggiani}, {Pellizzoni}, {Picozza}, {Pilia},
  {Rappoldi}, {Trois}, {Vercellone}, {Vittorini}, {ASAS-SN Team}, {Stanek},
  {Franckowiak}, {Kochanek}, {Beacom}, {Thompson}, {Holoien}, {Dong}, {Prieto},
  {Shappee}, {Holmbo}, {HAWC Collaboration}, {Abeysekara}, {Albert}, {Alfaro},
  {Alvarez}, {Arceo}, {Arteaga-Vel{\'a}zquez}, {Avila Rojas}, {Ayala Solares},
  {Becerril}, {Belmont-Moreno}, {Bernal}, {Caballero-Mora}, {Capistr{\'a}n},
  {Carrami{\~n}ana}, {Casanova}, {Castillo}, {Cotti}, {Cotzomi}, {Couti{\~n}o
  de Le{\'o}n}, {De Le{\'o}n}, {De la Fuente}, {Diaz Hernandez}, {Dichiara},
  {Dingus}, {DuVernois}, {D{\'\i}az-V{\'e}lez}, {Ellsworth}, {Engel},
  {Fiorino}, {Fleischhack}, {Fraija}, {Garc{\'\i}a-Gonz{\'a}lez}, {Garfias},
  {Gonz{\'a}lez Mu{\~n}oz}, {Gonz{\'a}lez}, {Goodman}, {Hampel-Arias},
  {Harding}, {Hernandez}, {Hona}, {Hueyotl-Zahuantitla}, {Hui},
  {H{\"u}ntemeyer}, {Iriarte}, {Jardin-Blicq}, {Joshi}, {Kaufmann}, {Kunde},
  {Lara}, {Lauer}, {Lee}, {Lennarz}, {Le{\'o}n Vargas}, {Linnemann},
  {Longinotti}, {Luis-Raya}, {Luna-Garc{\'\i}a}, {Malone}, {Marinelli},
  {Martinez}, {Martinez-Castellanos}, {Mart{\'\i}nez-Castro},
  {Mart{\'\i}nez-Huerta}, {Matthews}, {Miranda-Romagnoli}, {Moreno},
  {Mostaf{\'a}}, {Nayerhoda}, {Nellen}, {Newbold}, {Nisa}, {Noriega-Papaqui},
  {Pelayo}, {Pretz}, {P{\'e}rez-P{\'e}rez}, {Ren}, {Rho}, {Rivi{\`e}re},
  {Rosa-Gonz{\'a}lez}, {Rosenberg}, {Ruiz-Velasco}, {Ruiz-Velasco}, {Salesa
  Greus}, {Sandoval}, {Schneider}, {Schoorlemmer}, {Sinnis}, {Smith},
  {Springer}, {Surajbali}, {Tibolla}, {Tollefson}, {Torres}, {Villase{\~n}or},
  {Weisgarber}, {Werner}, {Yapici}, {Gaurang}, {Zepeda}, {Zhou}, {{\'A}lvarez},
  {H.~E.~S.~S. Collaboration}, {Abdalla}, {Ang{\"u}ner}, {Armand}, {Backes},
  {Becherini}, {Berge}, {B{\"o}ttcher}, {Boisson}, {Bolmont}, {Bonnefoy},
  {Bordas}, {Brun}, {B{\"u}chele}, {Bulik}, {Caroff}, {Carosi}, {Casanova},
  {Cerruti}, {Chakraborty}, {Chandra}, {Chen}, {Colafrancesco}, {Davids},
  {Deil}, {Devin}, {Djannati-Ata{\"\i}}, {Egberts}, {Emery}, {Eschbach},
  {Fiasson}, {Fontaine}, {Funk}, {F{\"u}{\ss}ling}, {Gallant}, {Gat{\'e}},
  {Giavitto}, {Glawion}, {Glicenstein}, {Gottschall}, {Grondin}, {Haupt},
  {Henri}, {Hinton}, {Hoischen}, {Holch}, {Huber}, {Jamrozy}, {Jankowsky},
  {Jankowsky}, {Jouvin}, {Jung-Richardt}, {Kerszberg}, {Kh{\'e}lifi}, {King},
  {Klepser}, {Kluz {\textasciiacute}niak}, {Komin}, {Kraus}, {Lefaucheur},
  {Lemi{\`e}re}, {Lemoine-Goumard}, {Lenain}, {Leser}, {Lohse},
  {L{\'o}pez-Coto}, {Lorentz}, {Lypova}, {Marandon}, {Guillem
  Mart{\'\i}-Devesa}, {Maurin}, {Mitchell}, {Moderski}, {Mohamed}, {Mohrmann},
  {Moulin}, {Murach}, {de Naurois}, {Niederwanger}, {Niemiec}, {Oakes},
  {O'Brien}, {Ohm}, {Ostrowski}, {Oya}, {Panter}, {Parsons}, {Perennes},
  {Piel}, {Pita}, {Poireau}, {Priyana Noel}, {Prokoph}, {P{\"u}hlhofer},
  {Quirrenbach}, {Raab}, {Rauth}, {Renaud}, {Rieger}, {Rinchiuso}, {Romoli},
  {Rowell}, {Rudak}, {Sasaki}, {Sanchez}, {Schlickeiser}, {Sch{\"u}ssler},
  {Schulz}, {Schwanke}, {Seglar-Arroyo}, {Shafi}, {Simoni}, {Sol}, {Stegmann},
  {Steppa}, {Tavernier}, {Taylor}, {Tiziani}, {Trichard}, {Tsirou}, {van
  Eldik}, {van Rensburg}, {van Soelen}, {Veh}, {Vincent}, {Voisin}, {Wagner},
  {Wagner}, {Wierzcholska}, {Zanin}, {Zdziarski}, {Zech}, {Ziegler}, {Zorn},
  {{\.Z}ywucka}, {INTEGRAL Team}, {Savchenko}, {Ferrigno}, {Bazzano}, {Diehl},
  {Kuulkers}, {Laurent}, {Mereghetti}, {Natalucci}, {Panessa}, {Rodi},
  {Ubertini}, {Kanata}, Teams, {Morokuma}, {Ohta}, {Tanaka}, {Mori},
  {Yamanaka}, {Kawabata}, {Utsumi}, {Nakaoka}, {Kawabata}, {Nagashima},
  {Yoshida}, {Matsuoka}, {Itoh}, {Kapteyn Team}, {Keel}, {Liverpool Telescope
  Team}, {Copperwheat}, {Steele}, {Swift/NuSTAR Team}, {Cenko}, {Cowen},
  {DeLaunay}, {Evans}, {Fox}, {Keivani}, {Kennea}, {Marshall}, {Osborne},
  {Santander}, {Tohuvavohu}, {Turley}, {VERITAS Collaboration}, {Abeysekara},
  {Archer}, {Benbow}, {Bird}, {Brill}, {Brose}, {Buchovecky}, {Buckley},
  {Bugaev}, {Christiansen}, {Connolly}, {Cui}, {Daniel}, {Errando}, {Falcone},
  {Feng}, {Finley}, {Fortson}, {Furniss}, {Gueta}, {H{\"u}tten}, {Hervet},
  {Hughes}, {Humensky}, {Johnson}, {Kaaret}, {Kar}, {Kelley-Hoskins},
  {Kertzman}, {Kieda}, {Krause}, {Krennrich}, {Kumar}, {Lang}, {Lin}, {Maier},
  {McArthur}, {Moriarty}, {Mukherjee}, {Nieto}, {O'Brien}, {Ong}, {Otte},
  {Park}, {Petrashyk}, {Pohl}, {Popkow}, {Pueschel}, {Quinn}, {Ragan},
  {Reynolds}, {Richards}, {Roache}, {Rulten}, {Sadeh}, {Santander}, {Scott},
  {Sembroski}, {Shahinyan}, {Sushch}, {Tr{\'e}panier}, {Tyler}, {Vassiliev},
  {Wakely}, {Weinstein}, {Wells}, {Wilcox}, {Wilhelm}, {Williams}, {Zitzer},
  {VLA/B Team}, {Tetarenko}, {Kimball}, {Miller-Jones}, \&
  {Sivakoff}}]{icecube2018-II}
{IceCube Collaboration}, {Aartsen}, M.~G., {Ackermann}, M., {et~al.} 2018,
  Science, 361, eaat1378, \dodoi{10.1126/science.aat1378}

\bibitem[{Kislat {et~al.}(2015)Kislat, Clark, Beilicke, \&
  Krawczynski}]{kislat_analyzing_2015}
Kislat, F., Clark, B., Beilicke, M., \& Krawczynski, H. 2015, Astroparticle
  Physics, 68, 45, \dodoi{10.1016/j.astropartphys.2015.02.007}

\bibitem[{Kochenderfer \& Wheeler(2019)}]{kochenderfer_algorithms_2019}
Kochenderfer, M.~J., \& Wheeler, T.~A. 2019, Algorithms for {Optimization}
  (Cambridge, MA, USA: MIT Press)

\bibitem[{{Liodakis} {et~al.}(2019){Liodakis}, {Peirson}, \&
  {Romani}}]{Liodakis2019}
{Liodakis}, I., {Peirson}, A.~L., \& {Romani}, R.~W. 2019, \apj, 880, 29,
  \dodoi{10.3847/1538-4357/ab2719}

\bibitem[{{Liodakis} {et~al.}(2018){Liodakis}, {Romani}, {Filippenko},
  {Kiehlmann}, {Max-Moerbeck}, {Readhead}, \& {Zheng}}]{Liodakis2018}
{Liodakis}, I., {Romani}, R.~W., {Filippenko}, A.~V., {et~al.} 2018, \mnras,
  480, 5517, \dodoi{10.1093/mnras/sty2264}

\bibitem[{{Marscher}(2014)}]{Marscher2014}
{Marscher}, A.~P. 2014, \apj, 780, 87, \dodoi{10.1088/0004-637X/780/1/87}

\bibitem[{Padovani {et~al.}(2019)Padovani, Oikonomou, Petropoulou, Giommi, \&
  Resconi}]{padovani_txs_2019}
Padovani, P., Oikonomou, F., Petropoulou, M., Giommi, P., \& Resconi, E. 2019,
  Monthly Notices of the Royal Astronomical Society: Letters, 484, L104,
  \dodoi{10.1093/mnrasl/slz011}

\bibitem[{Peirson(2022)}]{peirson_alpv95sscpol_2022}
Peirson, A. 2022, alpv95/{SSCpol}: v1.3.1,  Zenodo,
  \dodoi{10.5281/zenodo.6450411}

\bibitem[{{Peirson} \& {Romani}(2018)}]{Peirson2018}
{Peirson}, A.~L., \& {Romani}, R.~W. 2018, \apj, 864, 140,
  \dodoi{10.3847/1538-4357/aad69d}

\bibitem[{{Peirson} \& {Romani}(2019)}]{Peirson2019}
---. 2019, \apj, 885, 76, \dodoi{10.3847/1538-4357/ab46b1}

\bibitem[{Peirson \& Romani(2021)}]{peirson_towards_2021}
Peirson, A.~L., \& Romani, R.~W. 2021, arXiv:2107.08289 [astro-ph].
\newblock \url{http://arxiv.org/abs/2107.08289}

\bibitem[{Peirson {et~al.}(2021)Peirson, Romani, Marshall, Steiner, \&
  Baldini}]{peirson_deep_2021}
Peirson, A.~L., Romani, R.~W., Marshall, H.~L., Steiner, J.~F., \& Baldini, L.
  2021, Nuclear Instruments and Methods in Physics Research Section A:
  Accelerators, Spectrometers, Detectors and Associated Equipment, 986, 164740,
  \dodoi{10.1016/j.nima.2020.164740}

\bibitem[{Pesce-Rollins {et~al.}(2019)Pesce-Rollins, Lalla, Omodei, \&
  Baldini}]{pesce-rollins_observation-simulation_2019}
Pesce-Rollins, M., Lalla, N.~D., Omodei, N., \& Baldini, L. 2019, Nuclear
  Instruments and Methods in Physics Research A, 936, 224,
  \dodoi{10.1016/j.nima.2018.10.041}

\bibitem[{Potter \& Cotter(2012)}]{potter_synchrotron_2012}
Potter, W.~J., \& Cotter, G. 2012, Monthly Notices of the Royal Astronomical
  Society, 423, 756, \dodoi{10.1111/j.1365-2966.2012.20918.x}

\bibitem[{Rubinstein \& Kroese(2004)}]{rubinstein_cross_2004}
Rubinstein, R.~Y., \& Kroese, D.~P. 2004, The {Cross} {Entropy} {Method}: {A}
  {Unified} {Approach} {To} {Combinatorial} {Optimization}, {Monte}-carlo
  {Simulation} ({Information} {Science} and {Statistics}) (Berlin, Heidelberg:
  Springer-Verlag)

\bibitem[{Simmons \& Stewart(1985)}]{simmons_point_1985}
Simmons, J. F.~L., \& Stewart, B.~G. 1985, Astronomy and Astrophysics, 142,
  100.
\newblock \url{https://ui.adsabs.harvard.edu/abs/1985A&A...142..100S/abstract}

\bibitem[{{Tavecchio} {et~al.}(2018){Tavecchio}, {Landoni}, {Sironi}, \&
  {Coppi}}]{Tavecchio2018}
{Tavecchio}, F., {Landoni}, M., {Sironi}, L., \& {Coppi}, P. 2018, \mnras, 480,
  2872, \dodoi{10.1093/mnras/sty1491}

\bibitem[{{Weisskopf} {et~al.}(2021){Weisskopf}, {Soffitta}, {Baldini},
  {Ramsey}, {O'Dell}, {Romani}, {Matt}, {Deininger}, {Baumgartner},
  {Bellazzini}, {Costa}, {Kolodziejczak}, {Latronico}, {Marshall}, {Muleri},
  {Bongiorno}, {Tennant}, {Bucciantini}, {Dovciak}, {Marin}, {Marscher},
  {Poutanen}, {Slane}, {Turolla}, {Kalinowski}, {Di Marco}, {Fabiani},
  {Minuti}, {La Monaca}, {Pinchera}, {Rankin}, {Sgro'}, {Trois}, {Xie},
  {Alexander}, {Allen}, {Amici}, {Andersen}, {Antonelli}, {Antoniak},
  {Attina'}, {Barbanera}, {Bachetti}, {Baggett}, {Bladt}, {Brez}, {Bonino},
  {Boree}, {Borotto}, {Breeding}, {Brienza}, {Bygott}, {Caporale}, {Cardelli},
  {Carpentiero}, {Castellano}, {Castronuovo}, {Cavalli}, {Cavazzuti},
  {Ceccanti}, {Centrone}, {Citraro}, {D' Amico}, {D'Alba}, {Di Gesu}, {Del
  Monte}, {Dietz}, {Di Lalla}, {Di Persio}, {Dolan}, {Donnarumma},
  {Evangelista}, {Ferrant}, {Ferrazzoli}, {Ferrie}, {Footdale}, {Forsyth},
  {Foster}, {Garelick}, {Gunji}, {Gurnee}, {Head}, {Hibbard}, {Johnson},
  {Kelly}, {Kilaru}, {Lefevre}, {Le Roy}, {Loffredo}, {Lorenzi}, {Lucchesi},
  {Maddox}, {Magazzu}, {Maldera}, {Manfreda}, {Mangraviti}, {Marengo},
  {Marrocchesi}, {Massaro}, {Mauger}, {McCracken}, {McEachen}, {Mize}, {Mereu},
  {Mitchell}, {Mitsuishi}, {Morbidini}, {Mosti}, {Nasimi}, {Negri}, {Negro},
  {Nguyen}, {Nitschke}, {Nuti}, {Onizuka}, {Oppedisano}, {Orsini}, {Osborne},
  {Pacheco}, {Paggi}, {Painter}, {Pavelitz}, {Pentz}, {Piazzolla}, {Perri},
  {Pesce-Rollins}, {Peterson}, {Pilia}, {Profeti}, {Puccetti}, {Ranganathan},
  {Ratheesh}, {Reedy}, {Root}, {Rubini}, {Ruswick}, {Sanchez}, {Sarra},
  {Santoli}, {Scalise}, {Sciortino}, {Schroeder}, {Seek}, {Sosdian}, {Spandre},
  {Speegle}, {Tamagawa}, {Tardiola}, {Tobia}, {Thomas}, {Valerie}, {Vimercati},
  {Walden}, {Weddendorf}, {Wedmore}, {Welch}, {Zanetti}, \&
  {Zanetti}}]{Weisskopf2022}
{Weisskopf}, M.~C., {Soffitta}, P., {Baldini}, L., {et~al.} 2021, arXiv
  e-prints, arXiv:2112.01269.
\newblock \doarXiv{2112.01269}

\bibitem[{{Wierzcholska} \& {Siejkowski}(2016)}]{Wierzcholska2016}
{Wierzcholska}, A., \& {Siejkowski}, H. 2016, \mnras, 458, 2350,
  \dodoi{10.1093/mnras/stw425}

\bibitem[{Zhang \& Boettcher(2013)}]{zhang_x-ray_2013}
Zhang, H., \& Boettcher, M. 2013, The Astrophysical Journal, 774, 18,
  \dodoi{10.1088/0004-637X/774/1/18}

\bibitem[{Zhang {et~al.}(2019)Zhang, Fang, Li, Giannios, Böttcher, \&
  Buson}]{zhang_probing_2019}
Zhang, H., Fang, K., Li, H., {et~al.} 2019, The Astrophysical Journal, 876,
  109, \dodoi{10.3847/1538-4357/ab158d}

\bibitem[{Zhang {et~al.}(2020)Zhang, Li, Giannios, Guo, Liu, \&
  Dong}]{zhang_radiation_2020}
Zhang, H., Li, X., Giannios, D., {et~al.} 2020, The Astrophysical Journal, 901,
  149, \dodoi{10.3847/1538-4357/abb1b0}

\bibitem[{Zhang {et~al.}(2018)Zhang, Santangelo, Feroci, Xu, Lu, Chen, Feng,
  Zhang, Brandt, Hernanz, Baldini, Bozzo, Campana, Rosa, Dong, Evangelista,
  Karas, Meidinger, Meuris, \& Zwart}]{zhang_enhanced_2018}
Zhang, S., Santangelo, A., Feroci, M., {et~al.} 2018, The enhanced {X}-ray
  {Timing} and {Polarimetry} mission - {eXTP}, Vol.~62,
  \dodoi{10.1007/s11433-018-9309-2}

\end{thebibliography}
\bibliographystyle{aasjournal}

\end{document}